\pdfoutput=1

\documentclass[11pt,twoside,a4paper,cmspaper,final,collab]{cms-tdr}
\def\svnVersion{438458P}\def\cmsCernNoTag{CERN-EP-2017-194}\def\cmsCernDate{\today}\def\cmsMessage{Published in Physical Review Letters as \href{http://dx.doi.org/10.1103/PhysRevLett.119.221802}{\doi{10.1103/PhysRevLett.119.221802.}}}

\begin{document}\cmsNoteHeader{EXO-17-006}

\hyphenation{had-ron-i-za-tion}
\hyphenation{cal-or-i-me-ter}
\hyphenation{de-vices}
\RCS$Revision: 428464 $
\RCS$HeadURL: svn+ssh://svn.cern.ch/reps/tdr2/papers/EXO-17-006/trunk/EXO-17-006.tex $
\RCS$Id: EXO-17-006.tex 428464 2017-10-09 13:33:03Z alverson $

\newcommand{\fullLumi}{35.9\fbinv}
\newcommand{\mt}{\ensuremath{M_\mathrm{T}} }
\newcommand{\WZ}{\ensuremath{\PW\Z}\xspace}
\newcommand{\ZZ}{\ensuremath{\Z\Z}\xspace}
\newcommand{\LT}{\ensuremath{L_\mathrm{T}}\xspace}
\newcommand{\Szero}{\ensuremath{\varSigma^{0}}}
\newcommand{\ls}{\ensuremath{\ell}}
\newcommand{\lspm}{\ensuremath{\ell^{\pm}}}
\newcommand{\lsmp}{\ensuremath{\ell^{\mp}}}

\newlength\cmsFigWidth
\ifthenelse{\boolean{cms@external}}{\setlength\cmsFigWidth{0.85\columnwidth}}{\setlength\cmsFigWidth{0.4\textwidth}}
\ifthenelse{\boolean{cms@external}}{\providecommand{\cmsLeft}{top\xspace}}{\providecommand{\cmsLeft}{left\xspace}}
\ifthenelse{\boolean{cms@external}}{\providecommand{\cmsRight}{bottom\xspace}}{\providecommand{\cmsRight}{right\xspace}}
\ifthenelse{\boolean{cms@external}}{\providecommand{\NA}{\ensuremath{\cdots}}}{\providecommand{\NA}{\ensuremath{\text{---}}}}
\ifthenelse{\boolean{cms@external}}{\providecommand{\CL}{C.L.}}{\providecommand{\CL}{CL}}

\cmsNoteHeader{EXO-17-006}
\title{Search for evidence of the type-III seesaw mechanism in multilepton final states in proton--proton collisions at \texorpdfstring{$\sqrt{s} = 13\TeV$}{sqrt(s) = 13 TeV}}

\date{\today}

\abstract{
A search for a signal consistent with the type-III seesaw mechanism in events with three or more electrons or muons is presented.
The data sample consists of proton-proton collisions at $\sqrt{s} = 13\TeV$  collected by the CMS experiment at the LHC in 2016 and corresponds to an integrated luminosity of 35.9\fbinv.
Selection criteria based on the number of leptons and the invariant mass of oppositely charged lepton pairs are used to distinguish the signal from the standard model background.
The observations are consistent with the expectations from standard model processes.
The results are used to place limits on the production of heavy fermions of the type-III seesaw model as a function of the branching ratio to each lepton flavor.
In the scenario of equal branching fractions to each lepton flavor, heavy fermions with masses below 840\GeV are excluded. This is the most sensitive probe to date of the type-III seesaw mechanism.
}

\hypersetup{%
pdfauthor={CMS Collaboration},%
pdftitle={Search for evidence of type-III seesaw mechanism in multilepton final states in proton--proton collisions at sqrt(s) = 13 TeV},%
pdfsubject={CMS EXO Analyses},%
pdfkeywords={EXO}}

\maketitle

\label{sec:Introduction}

The discovery of neutrino oscillations established that neutrinos have mass~\cite{Nustatus}.
Among proposed extensions of the Standard Model (SM) that explain neutrino mass, models based on the seesaw mechanism
are an appealing possibility~\cite{typeIa, typeIb,typeIIa,typeIIe,seesawII_2,seesawII_6, SeesawIII_first, PhysRevLett.56.561,PhysRevD.34.1642}.
The seesaw mechanism introduces new heavy particles that couple to leptons and to a Higgs boson (\PH) doublet, and accounts
for the smallness of neutrino masses~\cite{Wang:2016lve, Li:2009mw, Deppisch:2015qwa}. Within the type-III seesaw
model~\cite{SeesawIII_first}, the neutrino is considered a Majorana particle whose mass arises from the mediation of new
massive fermions that form an  SU(2) triplet of heavy Dirac charged leptons $\PgS^{\pm}$and a
heavy Majorana neutral lepton $\PgSz$. In proton-proton ($\Pp\Pp$) collisions, these massive fermions may be pair-produced through
leading order electroweak interactions, in both charged-charged and charged-neutral pairs.
The renewed interest in the type-III seesaw models~\cite{FileviezPerez:2008sr, Arhrib:2009mz, delAguila:2008hw, Franceschini:2008pz, McDonald:2013hsa, Ding:2014nga} emphasizes the importance of exploring such signatures at the CERN LHC.

We conduct a search for these new massive fermions by examining final states with at least three electrons and muons
(including those that come from the decays of tau leptons)~\cite{delAguila:2008cj,delAguila:2009bb}. Such final states arise from these
decays: $\PgS^{\pm} \to \PW^{\pm}\nu_\ell$, $\PgS^{\pm} \to \PZ\lspm$, $\PgS^{\pm} \to \PH\lspm$,
$\PgSz\to\PW^{\pm}\lsmp$, $\PgSz \to \PZ\nu_\ell$, and $\PgSz\to\PH\nu_\ell$, where $\ls = \Pe, \Pgm$, or $\Pgt$.
The branching fractions are mass dependent. A complete decay chain example would be
$\PgS^{\pm}\PgSz\to\PW^{\pm}\Pgn\PW^{\pm}\lsmp \to \lspm\Pgn\Pgn\lspm\Pgn\lsmp$.
In this search, we consider all 27 different production and decay processes resulting from the
decays of each of the $\PgSz\PgSp$, $\PgSz\PgSm$, and $\PgSp\PgSm$ pairs to the nine different pairs of $\PW$, $\PZ$ and $\PH$ bosons.

In the model used in this analysis~\cite{SeesawIII_Biggio}, the $\PgS^{\pm}$ and $\PgSz$ decays are prompt, and gauge invariance ensures that they
are degenerate in mass at tree level.
The branching fraction ($B_{\ell}$) of a heavy fermion to a lepton of flavor
$\ls$ is proportional to $\abs{V_{\ell}}^2/(\abs{V_{\Pe}}^2+\abs{V_{\Pgm}}^2+\abs{V_{\Pgt}}^2)$, where $V_\ell$ is the heavy-light fermion mixing angle.
A prior CMS search using 7\TeV data excluded heavy fermions with mass ($m_{\varSigma}$)
below 179\GeV~\cite{CMS-PAPER-EXO-11-073}, probing a scenario where the
branching fractions to all lepton flavors are equal (flavor-democratic scenario, $B_{\Pe} =B_{\Pgm} =B_{\Pgt} $).
Using 8\TeV data, the ATLAS Collaboration similarly excluded $m_{\PgS}$ below 335\GeV,
assuming that the $\PgS$s decay only to the first two generations of leptons~\cite{ATLAS-CERN-PH-EP-2015-094}.
In this Letter, we probe all values of $B_{\ell}$ for each flavor of $\ell$, subject to the constraint $B_\Pe+B_\Pgm+B_\Pgt=1$.

We pursue a search in final states with at least three charged leptons. The backgrounds can be classified as the irreducible ones due to
the production and decay of dibosons (\WZ and \ZZ production), and the reducible ones due to the leptonic decay of $\Z+$jets or $\ttbar+$jets
accompanied by leptons originating from heavy quark decay, or from misidentification of jets as leptons. In addition, small irreducible contributions
to the background arise from other SM processes such as $\ttbar \PW$, $\ttbar \Z$, triboson ($\Z{}\Z{}\Z$, $\PW{}\PW{}\PZ$ etc.), and Higgs boson production.

\label{sec:detector}

The central feature of the CMS apparatus is a superconducting solenoid of 6\unit{m} internal diameter, providing a magnetic field of 3.8\unit{T}. Within the solenoid volume are a silicon pixel and strip tracker, a lead tungstate crystal electromagnetic calorimeter (ECAL), and a brass and scintillator hadron calorimeter, each composed of a barrel and two endcap sections. Forward calorimeters extend the pseudorapidity ($\eta$) coverage provided by the barrel and endcap detectors. Muons are detected in gas-ionization chambers embedded in the steel flux-return yoke outside the solenoid. A detailed description of the CMS detector can be found in Ref.~\cite{Chatrchyan:2008zzk}.

\label{sec:SSB}

The data sample consists of $\Pp\Pp$ collisions at $\sqrt{s} = 13\TeV$ collected by the CMS experiment at the LHC in 2016, corresponding to an
integrated luminosity of \fullLumi.  The data for this search are collected using several dilepton triggers based on a two-level trigger
system~\cite{Khachatryan:2016bia}. The double-electron (double-muon) trigger requires two loosely isolated electrons (muons). The
transverse momentum (\pt) thresholds on the leading (subleading) lepton are 23 (12)\GeV for the double-electron, and 17 (8)\GeV for
the double-muon trigger, respectively. We also use two muon/electron cross-trigger combinations that require a $\pt > 23\GeV$ loosely isolated
muon (electron) and a $\pt>8\GeV$ loosely isolated electron (muon).

The signal is simulated using \MGvATNLO v2.2.2~\cite{Alwall:2014hca} at leading order precision using the
{NNPDF30\_lo\_as\_0130\_nf\_4}~\cite{Ball:2014uwa} parton distribution functions (PDFs). The signal cross section is calculated at next-to-leading order
(NLO) plus next-to-leading logarithmic accuracy, assuming that the heavy leptons, $\PgS^{\pm,0}$, are {SU(2)} triplet
fermions \cite{Fuks:2012qx,Fuks:2013vua}. The diboson backgrounds are generated at NLO with \POWHEG 2.0~\cite{Nason:2004rx, Frixione:2007vw, Alioli:2010xd}. Backgrounds from Higgs boson production are generated using the \POWHEG and \textsc{jhugen} 6.2.8~\cite{Gao:2010qx, Bolognesi:2012mm, Anderson:2013afp, Gritsan:2016hjl} generators. Triboson backgrounds are generated using \MGvATNLO.
Bosonic decays, parton showering, fragmentation, and hadronization are simulated for all samples using \PYTHIA 8.212 ~\cite{Sjostrand:2006za, Sjostrand:2014zea}.

The response of the CMS detector is simulated using dedicated software based on the \GEANTfour toolkit~\cite{Agostinelli:2002hh}.
The presence of multiple $\Pp\Pp$ interactions in the same or adjacent bunch crossing (pileup)~\cite{CMS-PAS-JME-14-001} is incorporated by simulating additional
interactions that are both in-time and out-of-time with the collision.

\label{sec:reconstruction}

Events satisfying the trigger are required to pass additional offline selections. Electrons and muons must satisfy $\pt>10\GeV$ and $\abs{\eta}<2.5$
($\abs{\eta}<2.4$ for muons). Electrons are reconstructed using Gaussian Sum Filter tracks matched to ECAL superclusters~\cite{Khachatryan:2015hwa}.
Muons are reconstructed using the particle-flow (PF) algorithm~\cite{Sirunyan:2017ulk, CMS-PAS-PFT-09-001} that utilizes measurements from the
tracker, calorimeter and muon systems to reconstruct candidate muons, photons, and hadrons. Tracks of candidate electrons and
muons must satisfy quality requirements and should match geometrically with either energy deposits in the ECAL (for electrons) or tracks in the
muon detectors (for muons).

Jets are reconstructed from the PF candidates and electrons using the anti-\kt algorithm~\cite{Cacciari:2008gp}
with distance parameter $R=0.4$. Jets are required to have $\pt \geq 30\GeV$ and $\abs{\eta}\leq 3.0$ and satisfy quality criteria that
remove contributions from pileup. The missing transverse momentum vector \ptvecmiss is defined as the projection onto the plane perpendicular
to the beam axis of the negative vector sum of the momenta of all reconstructed PF objects in an event~\cite{Khachatryan:2014gga}.
Its magnitude is referred to as \ptmiss. Jet energy corrections are applied to all jets and also propagated to the calculation of $\ptvecmiss$.

The reducible backgrounds include leptons inside or near jets and leptons from heavy quark decays.
In addition hadrons that reach the muon detectors, or hadronic showers with large electromagnetic energy fractions can be misidentified as leptons.
Henceforth we refer to all leptons not arising
from vector boson decay (either directly or via an intermediate tau lepton) as misidentified leptons. Backgrounds from misidentified leptons are significantly reduced by an isolation requirement comparing
the $\pt$ of a lepton to the summed $\pt$ of particles in its neighborhood. We utilize relative isolation, which is defined as the scalar $\pt$ sum,
normalized to the lepton $\pt$, of photon and hadron PF candidates within a cone around the lepton. This relative isolation is required to be less
than 25\% within a cone of size $R=\sqrt{\smash[b]{(\Delta\eta)^2+(\Delta\phi)^2}}=0.4$ for muons, where $\phi$ is azimuthal angle in radians. For electrons,
the relative isolation within a cone of $R = 0.3$ must be less than 7\% (8\%) for the reconstructed energy deposits situated in the
ECAL barrel [$\abs{\eta} < 1.479$] (ECAL endcap [$\abs{\eta} > 1.479$]). Charged tracks returned by a jet finding algorithm~\cite{Cacciari:2008gp,Cacciari:2011ma}
and the corresponding
\ptmiss\ are associated with each reconstructed vertex. The vertex with the largest value of $\Sigma\pt^2$ over the charged tracks and the \ptmiss\
is taken to be the primary $\Pp\Pp$ interaction vertex. All isolation quantities are corrected for contributions from particles originating from
pileup vertices.
Electrons and muons must satisfy $\abs{d_z}< 0.1\cm$ and $\abs{d_{xy}}<0.05\cm$, where $d_z$ and $d_{xy}$ are the longitudinal and
transverse impact parameters of leptons with respect to the primary
vertex, respectively.  For electrons in the ECAL endcap, we relax these requirements to $\abs{d_z} < 0.2 \cm$ and $\abs{d_{xy}} < 0.1\cm$.

\label{sec:eventselection}

For the two leading leptons we apply offline $\pt$ thresholds of 25 and 15\GeV. Additional leptons must satisfy $\pt>10\GeV$.
With these thresholds, the trigger efficiency for trilepton events is close to 100\%.
Events containing an opposite-sign same-flavor (OSSF) lepton pair with mass below 12\GeV are vetoed to reduce background from low-mass resonances.
We reject trilepton events containing an OSSF pair with mass below $81\GeV$ when the trilepton mass is within a \PZ boson mass window ($91\pm 10\GeV$).
This suppresses events from the $\Z \to \ls\ls^{\ast} \to \ls\ls\gamma$ background process, where the
photon converts into two additional leptons, one of which is lost.

We classify the selected multilepton events into statistically independent search channels using the number of leptons and OSSF pairs,
and other kinematic quantities as described below. The number of OSSF pairs is counted using each lepton once. For example, $\Pgmp\Pgmp\Pem$
is tagged as OSSF0, while $\Pgmp\Pgmm\Pep\Pem$ is tagged as OSSF2. We further classify events as ``on-\Z'', ``below-\Z'', and ``above-\Z'', based on the
presence and mass of at least one OSSF pair relative to a \PZ boson mass window, defined as the range 81--101\GeV. In cases of ambiguity
(such as $\Pgmp\Pgmm\Pgmm$),
the OSSF pair with its mass closest to that of the \PZ boson takes precedence. We define $\LT$ as the scalar sum of all charged lepton $\pt$s, and we
use $\LT+\ptmiss$ as the main discriminating variable.
This choice allows us to maintain sensitivity to decay modes such as $\PgS^{\pm} \to \lspm \PZ \to \lspm \lspm \lsmp$ with large
charged-lepton momenta, as well as $\Szero \to \PH\Pgn \to \PW\PW\Pgn$, where charged lepton $\pt$s are lower but the neutrinos
contribute to $\ptmiss$. For events with OSSF1 and on-\Z, the sensitivity is improved by considering the transverse mass,
$\mt = (2\ptmiss \pt^{\ell}[1-\cos(\ptvecmiss,\ptvec^{\ell})])^{1/2}$
using the lepton that is not part of the OSSF pair.
Events are grouped in eight bins of width 150\GeV and 100\GeV for $\LT+\ptmiss$ and $\mt$, respectively. In each case the highest bin includes overflow events.

Altogether, six event categories are considered with eight bins of $\LT+\ptmiss$ in (i) three leptons, OSSF1, above-\Z; (ii) three leptons, OSSF1,
below-\Z; (iii) three leptons, OSSF0; (iv) four or more leptons, OSSF1; (v) four or more leptons, OSSF2; and eight bins of $\mt$ in
(vi) three leptons, OSSF1, on-\Z. The 48 mutually exclusive signal regions are summarized in Table~\ref{tab:sigreg}.

\begin{table}
 \centering
    \topcaption{ The signal regions summarized in terms of the number of leptons, the presence of an OSSF pair, and
    the kinematic variable used for signal discrimination. Each region is divided into eight bins in the kinematic variable, leading to 48 independent signal regions.
    Additional criteria based on $\ptmiss$ ensure that signal and control regions do not overlap.}
    \begin{scotch}{ l l l l}
      $N_{\text{leptons}}$&OSSF \& mass & Variable& $\ptmiss$ requirement\\ \hline
       3&OSSF1, on-\Z& $\mt$& $\ptmiss>100\GeV$\\[2ex]
      \multirow{3}{*}{3} &OSSF1, above-\Z& $\LT$+$\ptmiss$& no requirement \\
      &OSSF1, below-\Z&  $\LT$+$\ptmiss$& $\ptmiss>50\GeV$\\
      &OSSF0& $\LT$+$\ptmiss$& no requirement\\[2ex]
      \multirow{2}{*}{$\geq 4$} &OSSF1 & $\LT$+$\ptmiss$& no requirement\\
      &OSSF2 & $\LT$+$\ptmiss$& $\ptmiss>50\GeV$ if on-\Z\\
      \end{scotch}
    \label{tab:sigreg}
  \end{table}

\label{sec:bkg}

The irreducible backgrounds are estimated using simulation samples. The dominant processes are \WZ and \ZZ production and are normalized to
data using dedicated control regions. The \WZ control sample consists of events with exactly three leptons, tagged as OSSF1 on-\Z, satisfying
$50< \ptmiss< 100\GeV$ and $\mt > 30\GeV$. The ratio of the predicted \WZ rate to data (after correcting for backgrounds other than \WZ) is
$1.15 \pm 0.08$, where the uncertainty includes statistical and systematic contributions. The \ZZ control sample consists of events with exactly
four leptons, OSSF2, on-\Z for both pairs, and with $\ptmiss< 50\GeV$. The ratio of the predicted \ZZ rate to data is $1.25 \pm 0.06$. These factors
are used to normalize the \WZ and \ZZ background estimates. We also use these control regions to verify the modeling of kinematic distributions
such as $\LT+\ptmiss$ and $\mt$. The ``rare'' irreducible backgrounds from triboson, $\ttbar \PW$, $\ttbar \PZ$, and Higgs boson processes are estimated
from simulated samples using the theoretical cross sections.

The misidentified lepton background is estimated from observations using a 3-dimensional implementation of a matrix
method~\cite{Chatrchyan:2012bra} in which probabilities (rates) of prompt and misidentified leptons satisfying a loose lepton
selection to also pass a tight lepton selection are measured in data using a dilepton selection for prompt leptons,
and a trilepton signal-depleted selection (OSSF1, on-\Z, $\ptmiss<50\GeV$) for misidentified leptons. The rates are parametrized as a function of
lepton $\pt$, $\eta$ and the number of PF candidates around the lepton. The rate measurements are dominated by $\PZ$+jets events, and are corrected
using simulation to an average of $\PZ$+jets and $\ttbar$+jets events.

A residual ``conversion'' background is also estimated using the data. When a final-state radiation photon from a lepton converts
asymmetrically, often only one of the resultant leptons is reconstructed in the detector.
To estimate this background, we count
trilepton events where the trilepton mass is on-\Z, but the dilepton mass is below-\Z. These events are compared with dilepton+photon events
where the dilepton invariant mass is below-\Z, but the combined invariant mass of the two leptons and the photon is on-\Z. Photons suitable
for modeling conversions are selected by ignoring photons that are far away from a lepton or are close to a jet. The photon conversion rates are
3\% to 5\% depending upon the lepton type and $\pt$.
The conversion background in a signal region is estimated by scaling the observed number of dilepton+photon events with these rates.

\begin{figure}
\centering
\includegraphics[width=0.49\textwidth]{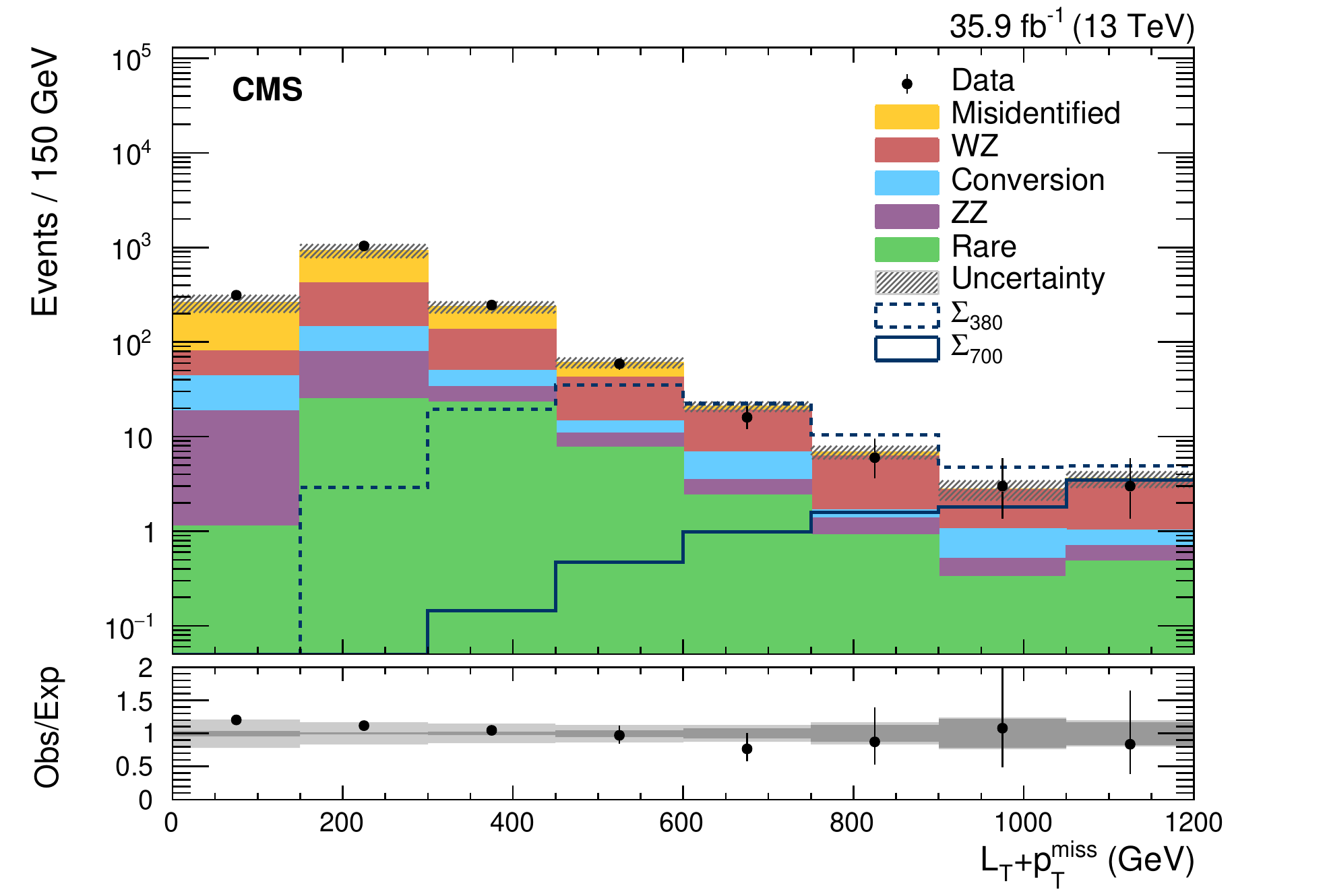}
\includegraphics[width=0.49\textwidth]{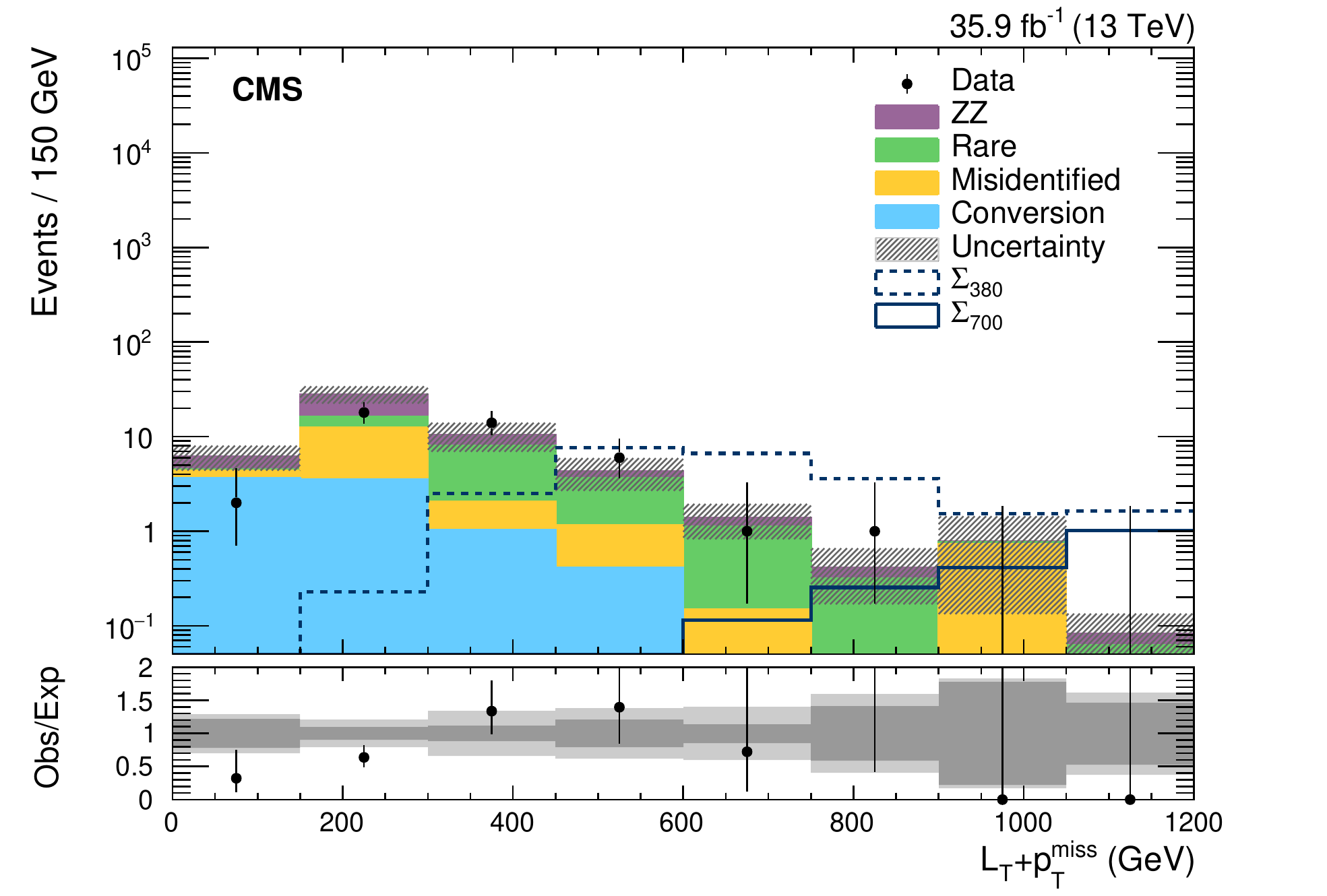}
\caption{
The $\LT+\ptmiss$  distribution for events with three leptons and one OSSF pair with mass
above-\Z (\cmsLeft) and the $\LT+\ptmiss$  distribution for events with four or more leptons and one
OSSF pair (\cmsRight). The total SM background is shown as a stacked histogram of all contributing
processes. The predictions for signal models (sum of all production and decay modes) with  $m_{\PgS} = 700\GeV$ (solid line) and $m_{\PgS} = 380\GeV$
(dashed line) are also shown. The lower panels show the ratio of observed to expected events. The hatched gray band in the upper panel, and the dark and light gray bands in the lower panel represent the total, statistical, and systematic uncertainties in the expected background, respectively.
\label{fig:sr}}
\end{figure}

\label{sec:systematic}

As an example of the complete background estimation, Fig.~\ref{fig:sr} shows the background predictions
and the observed events with three leptons, OSSF1, above-\Z, and for events with four or more leptons, OSSF1, along with systematic uncertainties.
The \WZ (\ZZ) background has a systematic uncertainty of 7\% (5\%) arising from the normalization factor measurement. The rare backgrounds have an
uncertainty of 50\% in the theoretical cross section and 2.5\% due to the integrated luminosity measurement. The misidentified background has an
uncertainty of 30\%, which is derived by varying the measured rates up and down for each lepton flavor within their respective uncertainties. This
uncertainty is dominated by the differences in rates in $\PZ$+jets and $\ttbar$+jets events. The conversion background has an uncertainty of 30\%
arising from contamination by nonconversion backgrounds in the control sample used to measure the conversion rate.

Corrections are applied to the simulated background events to account for differences from data in lepton identification and isolation efficiencies,
dilepton trigger efficiencies, as well as in the overall scale and resolution of energy and momentum measurements for electrons, muons, and jets.
The minor uncertainties due to these corrections are considered separately and result in a 1--10\% variation of the simulation-based background
yields across all signal regions.

The uncertainties in the signal acceptance due to the PDF uncertainties are estimated by utilizing the replica weights of the
{NNPDF30\_lo\_as\_0130\_nf\_4} PDF set~\cite{Butterworth:2015oua}, and are found to be less than 3\% in all signal bins. The signal yields also
have uncertainties due to the integrated luminosity measurement (2.5\%) and the the corrections applied to the simulation samples mentioned in
the previous paragraph.

\label{sec:result}

\begin{figure}
\centering
\includegraphics[width=0.5\textwidth]{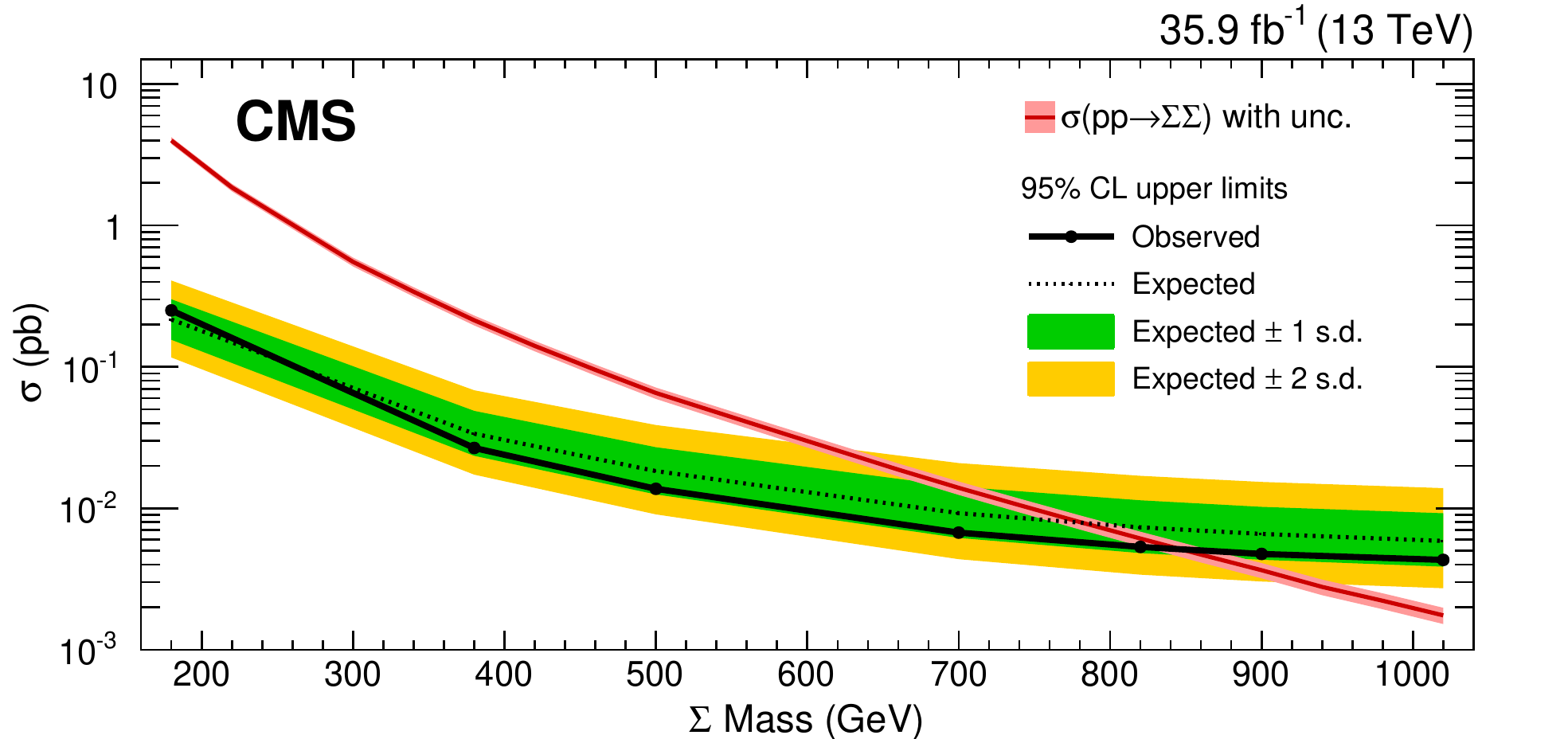}
\caption{\label{fig:limit} The 95\% confidence level upper limits on the cross section for production of heavy fermion pairs ($\PgSz \PgSp$, $\PgSz \PgSm$, and $\PgSp \PgSm$). Also shown is the theoretical prediction for the cross section of the \PgS{} pair production via the type III seesaw mechanism, with its uncertainty. In the flavor-democratic scenario ($B_\Pe=B_\Pgm=B_\Pgt$), heavy fermion pair production is excluded for masses below  840\GeV. }
\end{figure}

\begin{figure}
\centering
\includegraphics[width=0.49\textwidth]{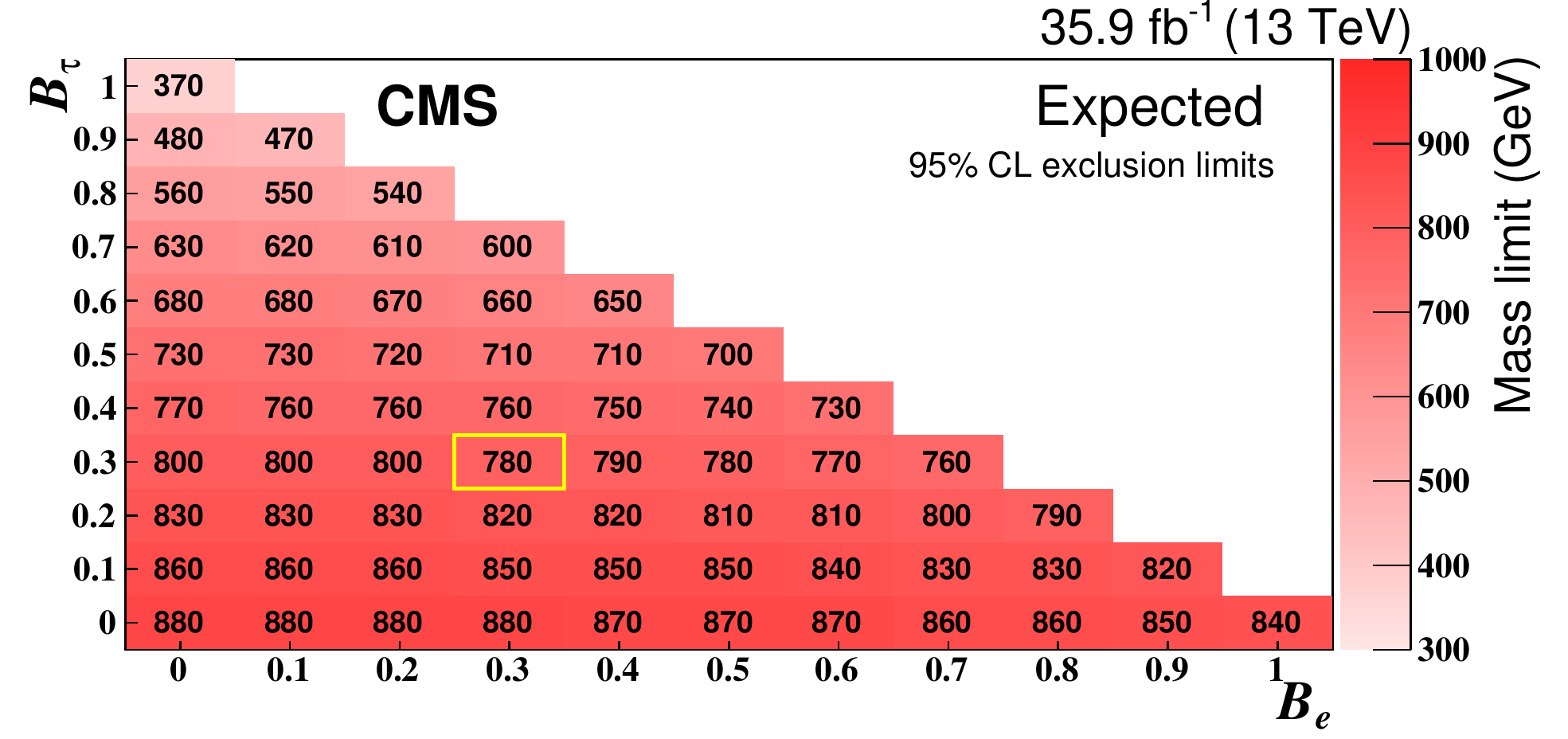}
\includegraphics[width=0.49\textwidth]{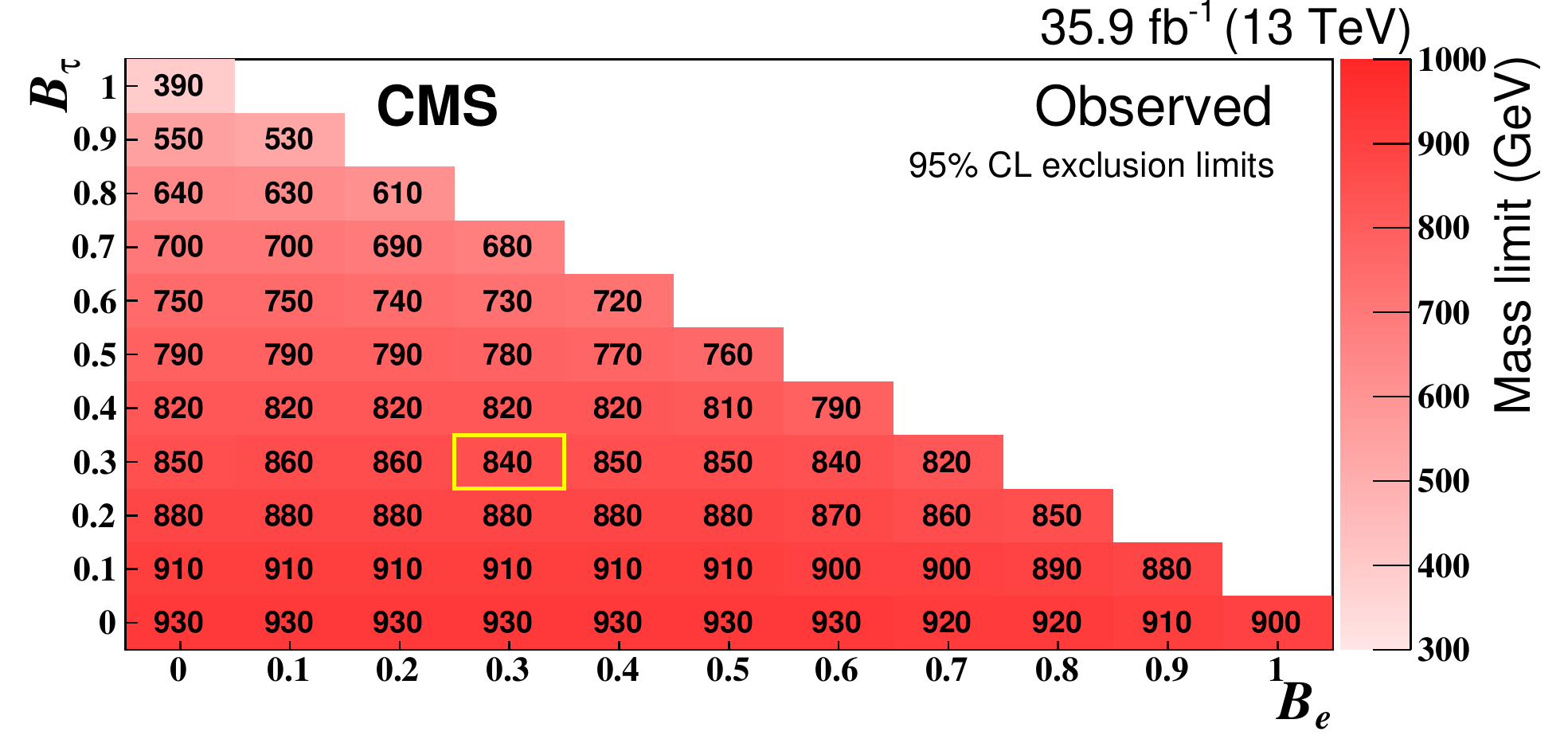}
\caption{The 95\% confidence level expected (\cmsLeft) and observed (\cmsRight) limits on the heavy fermion mass ${m_\PgS}$ (\GeV) for each combination of branching fractions to the individual lepton flavors. The color scale represents the mass exclusion limit obtained at each point. The branching fraction to $\Pgm$ can be obtained from $B_{\Pgm}=1-[B_{\Pe} + B_{\Pgt}]$. The highlighted box indicates the limit on $m_{\PgS}$ for the flavor-democratic scenario ($B_\Pe=B_\Pgm=B_\Pgt$).
\label{fig:limit2d}}
\end{figure}

We observe no statistically significant excess in the various signal regions that we probe.
We calculate a 95\% confidence level (\CL) upper limit on the sum of cross sections for
the production of heavy fermion pairs ($\PgSz \PgSp$, $\PgSz \PgSm$, and $\PgSp \PgSm$) using
the CL$_\mathrm{s}$ criterion~\cite{Junk:1999kv, Read:2002hq, CMS-NOTE-2011-005, Cowan:2010js}.
We consider all sources of systematic and statistical uncertainties as nuisance parameters, with log-normal and gamma prior distributions,
respectively, and perform a simultaneous fit across all search regions.
Figure~\ref{fig:limit} shows the upper limits on the cross section in the flavor-democratic scenario where the observed (expected) limit on masses of
heavy fermions is 840 (780)\GeV. The signal cross section has an uncertainty of approximately 5--15\% due to choices of PDFs, and factorization and
renormalization scales in the mass range we consider. Figure~\ref{fig:limit2d} shows the expected and observed excluded mass limits in the
$B_{\Pe}$--$B_{\Pgt}$ plane (note that $B_{\Pgm}=1-[B_{\Pe} + B_{\Pgt}]$).
The limits are the most restrictive when the $\PgS$ branching fraction to either electron or muon flavor is maximal, and become less restrictive
as the branching fraction to $\Pgt$ flavor increases.

\label{sec:conclusion}

In summary, a search has been performed for type-III seesaw heavy fermions in multilepton final states using \fullLumi of $\Pp\Pp$ collision data
at $\sqrt{s} = 13 \TeV$, collected using the CMS detector. No significant discrepancies are observed between the data
and the standard model prediction. Assuming degenerate heavy fermion masses,
previously unexplored regions of the signal model as a function of the branching ratio of heavy fermions to each lepton flavor are excluded.
In the lepton-flavor democratic scenario ($B_\Pe=B_\Pgm=B_\Pgt$), heavy fermions with mass below
840\GeV are excluded at 95\% confidence level. In the $\Pgt$-phobic case ($B_\Pgt=0$, $B_\Pe+B_\Pgm=1$) the mass
limits range from 900\GeV in the pure electron scenario to 930\GeV in the pure muon scenario. In the \Pe/\Pgm-phobic case ($B_\Pgt=1$, $B_\Pe+B_\Pgm=0$) the mass limit is 390\GeV. These are the strongest constraints to date
on the mass of heavy fermions associated with the type-III seesaw mechanism.

\begin{acknowledgments}
We congratulate our colleagues in the CERN accelerator departments for the excellent performance of the LHC and thank the technical and administrative staffs at CERN and at other CMS institutes for their contributions to the success of the CMS effort. In addition, we gratefully acknowledge the computing centers and personnel of the Worldwide LHC Computing Grid for delivering so effectively the computing infrastructure essential to our analyses. Finally, we acknowledge the enduring support for the construction and operation of the LHC and the CMS detector provided by the following funding agencies: BMWFW and FWF (Austria); FNRS and FWO (Belgium); CNPq, CAPES, FAPERJ, and FAPESP (Brazil); MES (Bulgaria); CERN; CAS, MoST, and NSFC (China); COLCIENCIAS (Colombia); MSES and CSF (Croatia); RPF (Cyprus); SENESCYT (Ecuador); MoER, ERC IUT, and ERDF (Estonia); Academy of Finland, MEC, and HIP (Finland); CEA and CNRS/IN2P3 (France); BMBF, DFG, and HGF (Germany); GSRT (Greece); OTKA and NIH (Hungary); DAE and DST (India); IPM (Iran); SFI (Ireland); INFN (Italy); MSIP and NRF (Republic of Korea); LAS (Lithuania); MOE and UM (Malaysia); BUAP, CINVESTAV, CONACYT, LNS, SEP, and UASLP-FAI (Mexico); MBIE (New Zealand); PAEC (Pakistan); MSHE and NSC (Poland); FCT (Portugal); JINR (Dubna); MON, RosAtom, RAS, RFBR and RAEP (Russia); MESTD (Serbia); SEIDI, CPAN, PCTI and FEDER (Spain); Swiss Funding Agencies (Switzerland); MST (Taipei); ThEPCenter, IPST, STAR, and NSTDA (Thailand); TUBITAK and TAEK (Turkey); NASU and SFFR (Ukraine); STFC (United Kingdom); DOE and NSF (USA). \end{acknowledgments}

\bibliography{auto_generated}

\cleardoublepage \appendix\section{The CMS Collaboration \label{app:collab}}\begin{sloppypar}\hyphenpenalty=5000\widowpenalty=500\clubpenalty=5000\textbf{Yerevan Physics Institute,  Yerevan,  Armenia}\\*[0pt]
A.M.~Sirunyan, A.~Tumasyan
\vskip\cmsinstskip
\textbf{Institut f\"{u}r Hochenergiephysik,  Wien,  Austria}\\*[0pt]
W.~Adam, F.~Ambrogi, E.~Asilar, T.~Bergauer, J.~Brandstetter, E.~Brondolin, M.~Dragicevic, J.~Er\"{o}, M.~Flechl, M.~Friedl, R.~Fr\"{u}hwirth\cmsAuthorMark{1}, V.M.~Ghete, J.~Grossmann, J.~Hrubec, M.~Jeitler\cmsAuthorMark{1}, A.~K\"{o}nig, N.~Krammer, I.~Kr\"{a}tschmer, D.~Liko, T.~Madlener, I.~Mikulec, E.~Pree, D.~Rabady, N.~Rad, H.~Rohringer, J.~Schieck\cmsAuthorMark{1}, R.~Sch\"{o}fbeck, M.~Spanring, D.~Spitzbart, W.~Waltenberger, J.~Wittmann, C.-E.~Wulz\cmsAuthorMark{1}, M.~Zarucki
% ...
\vskip\cmsinstskip
% ...
1:~~Also at Vienna University of Technology, Vienna, Austria\\
% ...

\end{sloppypar}
\end{document}